\documentclass[12pt]{article}
\usepackage{epsfig}
\usepackage{amsmath}

\normalsize

\def\bk{{\bf k}}

\def\bp{{\bf p}}
\def\bq{{\bf q}}

\def\b0{{\bf 0}}

\def\Im{{\rm Im}}

\def\alf{\alpha}
\def\eps{\epsilon}

\def\Gam{\Gamma}

\def\om{\omega}
\def\sg{\sigma}
\def\Sg{\Sigma}

\begin{document}


\title{\Large Soft Fermi Surfaces and Breakdown of Fermi Liquid Behavior}

\author{W.\ Metzner, D.\ Rohe, and S.\ Andergassen \\
{\small\em Max-Planck-Institut f\"ur Festk\"orperforschung, D-70569 Stuttgart, 
 Germany}}

\date{\small\today}
\maketitle


\begin{abstract}
Electron-electron interactions can induce Fermi surface 
deformations which break the point-group symmetry of the
lattice structure of the system.
In the vicinity of such a ''Pomeranchuk instability'' the 
Fermi surface is easily deformed by anisotropic perturbations, 
and exhibits enhanced collective fluctuations.
We show that critical Fermi surface fluctuations near a
d-wave Pomeranchuk instability in two dimensions lead to 
large anisotropic decay rates for single-particle excitations, 
which destroy Fermi liquid behavior over the whole surface 
except at the Brillouin zone diagonal. 
\noindent
\mbox{PACS: 71.10.Fd, 74.20.Mn} \\
\end{abstract}

Recently a number of authors pointed out that the Coulomb 
interaction between electrons in a metal can lead to Fermi 
surface deformations which break the orientational symmetry 
of the system.
Referring to a stability criterion for normal Fermi liquids
by Pomeranchuk,\cite{Pom} we use the term {\em ''Pomeranchuk 
instability''} (PI) for such symmetry-breaking Fermi surface
shifts.
Deformations of $d_{x^2-y^2}$-type, for which the Fermi surface 
expands along the $k_x$-axis and shrinks along the $k_y$-axis 
(or vice versa), have been found for various model Hamiltonians 
on a two-dimensional square lattice: t-J,\cite{YK1}
Hubbard,\cite{HM,GKW,NM} and extended Hubbard model.\cite{VV}
For a Fermi level close to the van Hove singularity the 
PI occurs already in the weak-coupling 
regime, where perturbative methods may be applied.
Symmetry-broken Fermi surfaces in fully isotropic (not lattice) 
two- and three-dimensional Fermi liquids have also been 
considered.\cite{OKF}
From a pure symmetry-group point of view a PI 
leads to a ''nematic'' electron liquid
as defined by Kivelson et al.\cite{KFE} in their discussion of
a possible correspondence between electron states in doped Mott 
insulators and liquid crystal phases. 

Symmetry-breaking Fermi surface deformations generally compete 
with other instabilities, but may also coexist with other types 
of symmetry-breaking order. For example, a superconducting state
with a d-wave deformed Fermi surface \cite{fn1} is stabilized in 
the two-dimensional Hubbard model with a sizable next-to-nearest
neighbor hopping amplitude and an electron density near van Hove 
filling, at least at weak coupling.\cite{NM} 
Superconducting nematic states have also been considered as one 
among several possibilities in a general symmetry classification 
by Vojta et al.\cite{VZS}
In the following we will however focus on symmetry-breaking Fermi 
surface deformations in an otherwise {\em normal} state.

For electrons on a lattice, the PI breaks only a discrete symmetry, 
the point-group symmetry of the lattice.
Hence, no Goldstone mode exists and symmetry-broken states can
exist also at finite temperature in $d \ge 2$ dimensions.
Order parameter fluctuations thus suppress the PI
much less than competing instabilities toward states
which would break a continuous symmetry.
In the ground state a PI can be driven as a function of the 
electron density or other control parameters, leading to a quantum 
critical point at the transition between the symmetric and 
symmetry-broken states, provided that no first order transition
occurs.

An interacting electron system in the vicinity of a PI is 
characterized by a {\em ''soft''} Fermi surface, which can be 
deformed very easily, that is at low energy cost.
In this article we show that strong dynamical fluctuations
of such a soft Fermi surface affect physical properties of the
system very strongly, and lead in particular to a very fast 
decay of single-particle excitations. 
The anisotropy of the Fermi surface fluctuations in a lattice
system leads to a pronounced anisotropy of the single-particle
decay rate. For $d_{x^2-y^2}$-wave fluctuations the decay is 
maximal near the $k_x$- and $k_y$-axes and minimal near the
diagonal of the Brillouin zone.

The ''softness'' of a Fermi surface can be quantified by the 
Fermi surface susceptibility \cite{HM}
\begin{equation}
 \kappa_{\bk_F\bk'_F} = 
 \frac{\delta s_{\bk_F}}{\delta\mu_{\bk'_F}} \; ,
\end{equation}
which measures the Fermi surface shifts $\delta s_{\bk_F}$
for small momentum dependent shifts of the chemical potential
$\delta\mu_{\bk'_F}$ at points $\bk'_F$ on the Fermi surface.
Close to a PI one of the eigenvalues of the 
matrix $\kappa_{\bk_F\bk'_F}$ diverges, and the corresponding
eigenvector describes the shape of the incipient Fermi surface
deformation. 

It is intuitively plausible that critical Fermi surface 
fluctuations will strongly affect single-particle excitations and 
as a consequence the low energy properties of the system. 
To explore the resulting physics, we define and analyze a
phenomenological lattice model with an effective interaction 
chosen such that a PI occurs, but no other instabilities.
The model Hamiltonian reads
\begin{equation}
 H = \sum_{\bk,\sg} \eps_{\bk} \, n_{\bk\sg} +
 \frac{1}{2V} \sum_{\bk,\bk',\bq} f_{\bk\bk'}(\bq) \,
 n_{\bk}(\bq) \, n_{\bk'}(-\bq) \; ,
\end{equation}
where $\eps_{\bk}$ is a single-particle dispersion,
$n_{\bk}(\bq) = 
 \sum_{\sg} c^{\dag}_{\bk-\bq/2,\sg} c_{\bk+\bq/2,\sg}$, and
$V$ the volume of the system.
Since the PI is driven by interactions with
small momentum transfers, that is forward scattering, we choose 
a coupling function $f_{\bk\bk'}(\bq)$ which contributes only 
for relatively small momenta $\bq$. This suppresses other 
instabilities such as superconductivity or density waves. 
We emphasize that this model is adequate only if Fermi 
surface fluctuations are the dominant fluctuations in the system.
Otherwise it would have to be supplemented by other interactions
with large momentum transfers. The interplay of Fermi surface
and other fluctuations opens a wide field for investigations in
the future.

For an analytical treatment we assume that the momentum dependence 
of the coupling function in (2) is separable, that is
\begin{equation}
 f_{\bk\bk'}(\bq) = g(\bq) \, d_{\bk} \, d_{\bk'} \; .
\end{equation}
Although the above model can be defined in any dimension, we now 
focus on the particularly interesting case of a two-dimensional
system on a square lattice.
To generate a PI with $d_{x^2-y^2}$ 
symmetry, the form factors $d_{\bk}$ must have that symmetry,
such as $d_{\bk} = \cos k_x - \cos k_y$, and $g(\bq)$
has to be negative, at least for $\bq \to \b0$. 
The resulting Landau function $f_{\bk\bk'} = f_{\bk\bk'}(\b0)$
captures qualitatively the most pronounced features of
the Landau function obtained from renormalization group
calculations\cite{HM} and perturbation theory\cite{FHR} for the 
two-dimensional Hubbard model near van Hove filling. 
In particular, it is repulsive for momenta near two different 
van Hove points and attractive for momenta near a common one.

We compute the two-particle vertex function $\Gam$ from $f$ by
summing the series of bubble chains sketched in Fig.\ 1. 
By virtue of the separable structure of the interaction, the
series can be summed algebraically yielding
\begin{equation}
 \Gam_{\bk\bk'}(\bq,\om) =
 \frac{g(\bq)}{1 - g(\bq) \, \Pi_d(\bq,\om)} \, 
 d_{\bk} \, d_{\bk'} \; ,
\end{equation}
where $\Pi_d$ is the particle-hole bubble with a form factor
$d_{\bk}$ at the vertices, that is
\begin{equation}
 \Pi_d(\bq,\om) = - \int \frac{d^2p}{(2\pi)^2} \,
 \frac{f(\eps_{\bp+\bq/2}) - f(\eps_{\bp-\bq/2})}
 {\om - (\eps_{\bp+\bq/2} - \eps_{\bp-\bq/2}) + i0^+} \; 
 d_{\bp}^2 \; .
\end{equation}
Here $f$ is the Fermi function.
For small $\bq$ and $\om$ the d-wave bubble has the following
asymptotic behavior: 
$\Pi_d(\bq,0) = - N_d^F + a(\hat\bq) \, q^2 + {\cal O}(q^4)$ and
$\Im\Pi_d(\bq,\om) \to - c(\hat\bq) \, \om/q \,$ for $q,\om \to 0$
and $\om/q \to 0$,
where $N_d^F > 0$ is a weighted density of states, with each 
state at the Fermi level weighted by the squared form factor 
$d_{\bk_F}^2$, while $a(\hat\bq)$ and $c(\hat\bq)$ are real
coefficients depending only on the direction $\hat\bq$ of $\bq$,
but not on its length. The coefficient $c(\hat\bq)$ is always 
positive, but the sign of $a(\hat\bq)$ depends on $\hat\bq$ and
on the choice of $\eps_{\bk}$, $d_{\bk}$ and $\mu$. 

The PI sets in when the denominator of
Eq.\ (4) vanishes at zero frequency and vanishing momenta, 
that is 
$\lim_{\bq \to 0} \, g(\bq) \, \Pi_d(\bq,0) = 
 - g(\b0) \, N_d^F = 1$ 
at the critical point.
We assume that the $\bq$-dependence of $g(\bq)$ is such that
$g(\bq) \, \Pi_d(\bq,0) < 1$ for $\bq \neq 0$ at that point.
Otherwise an instability with a finite $\bq$-vector would set
in first. For a small $\bq$-vector this would lead to a phase
where the symmetry-breaking Fermi surface deformations are
slowly modulated across the system. We leave this case for
future studies.

At the critical point the two-particle vertex has the following
asymptotic form for small $\bq$, $\om$ and small $\om/q$:
\begin{equation}
 \Gam_{\bk\bk'}(\bq,\om) \sim \frac{d_{\bk} \, d_{\bk'}}
 {i c(\hat\bq) \, \om/q - \alf(\hat\bq) \, q^2}
\end{equation}
with $\alf(\hat\bq) = a(\hat\bq) - N_d^F \, g''(\hat\bq)/g(\b0)$, 
where $g''(\hat\bq)$ is the second derivative of $g(\bq)$ with 
respect to $q$ for $q \to 0$.
The denominator of the vertex has the same form as for other 
familiar critical points in metals, namely at the boundary to 
ferromagnetic \cite{Her,Mil} or phase separated \cite{CDG} 
states. 
Peculiar to the d-wave PI are the form factors in the numerator.

We stress that the PI obeys the symmetry conditions for a continuous 
phase transition, since odd powers of the order parameter are excluded
by symmetry from the corresponding Landau theory, in contrast 
to the case of phase separation, where cubic terms usually drive a 
first order transition (except for special cases where the cubic
term is tuned to zero).
Note also that the Pomeranchuk instability is not inhibited by
long-range Coulomb forces, since volume conserving Fermi surface
deformations do not generate charge inhomogeneities.

To estimate the decay rate for single-particle excitations in
the presence of critical Fermi surface fluctuations near the
quantum critical point, we have
computed the imaginary part of the self-energy to first order
in $\Gam_{\bk\bk'}(\bq,\om)$, evaluating the Feynman diagram
in Fig.\ 2. For momenta on the Fermi surface, the result is
\begin{equation}
 \Im\Sg(\bk_F,\eps) \propto d_{\bk_F}^2 \, \eps^{2/3}
\end{equation}
for small energies $\eps$ at $T=0$ and
\begin{equation}
 \Im\Sg(\bk_F,0) \propto d_{\bk_F}^2 \, T^{2/3}
\end{equation}
at low finite temperatures. Note that the integral leading
to the above result is dominated by small momentum transfers
$\bq$ and $\om \ll v_{\bk_F} q$, thus justifying the
asymptotic expansion of the vertex function. Furthermore, the 
dominant contributions to $\Im\Sg$ at the Fermi vector $\bk_F$ 
come from (small) $\bq$-vectors that are {\em tangential} to 
the Fermi surface at $\bk_F$. 

Not unexpectedly, the decay rate has the same energy dependence
as for the quantum critical point near phase separation in two 
dimensions.\cite{CDG}
Different is however the d-wave form factor making the decay 
rate strongly anisotropic. 
The decay rate is strongest near the van Hove points, while the 
leading terms vanish on the diagonal of the Brillouin zone.
Subleading terms will produce at least conventional Fermi liquid 
decay rates ($T^2 \log T$) on the diagonal, but faster decay
(intermediate between $T^{2/3}$ and $T^2 \log T$)
may be obtained due to higher order processes and interactions
with large momentum transfers, which couple different parts of 
the Fermi surface.

The singular self-energy in the critical region will drastically
modify the one-particle propagator at low energies. However, 
this does not invalidate the calculation of the two-particle
vertex from bubbles with bare propagators, since singular 
self-energy corrections generated by strong forward scattering 
are cancelled by corresponding vertex corrections in the
polarization bubble.\cite{MCD} 
A more subtle point is whether higher order corrections to the
self-energy contribution in Fig.\ 2 modify the power-laws (7) and 
(8) for the decay rate. The same question has been discussed at 
length for fermions coupled to a gauge field,\cite{MCD}
where the lowest order calculation also yields a power-law with
exponent $2/3$.\cite{Lee}
A detailed analysis by Altshuler et al.\ \cite{AIM} indicated 
that this result is not changed by other terms.
A recent renormalization group analysis further supported the
validity of the leading power-law for the gauge theory, and
also for the critical point near phase separation in two
dimensions.\cite{CCDM}
Even if the above power-law would be modified by higher order
corrections, it is clear that Fermi liquid behavior cannot be 
restored, and the decay rate will remain large and anisotropic 
in any case.

A strongly anisotropic decay rate following a power-law with
exponent $2/3$ has recently been derived for an {\em isotropic}
continuum (not lattice) version of model (2).\cite{OKF} 
That result was obtained for the symmetry-broken ''nematic'' 
phase, and the anisotropy of the decay rate arises from 
the anisotropy of the symmetry-broken state and its collective
modes. In the symmetric phase of an isotropic liquid the decay 
rates are of course isotropic. At the quantum critical point
the decay rate of the isotropic liquid also obeys a power law
with exponent 2/3, but now isotropically over the whole Fermi 
surface.

Could soft Fermi surfaces and critical Fermi surface fluctuations 
play a role in cuprate superconductors? 
Due to the coupling of electron and lattice degrees of freedom
a symmetry-breaking Fermi surface deformation is generally 
accompanied by a lattice distortion, and vice versa. 
Structural transitions which reduce the lattice symmetry of 
the cuprate-planes are quite frequent in cuprates.
Close to a PI of the electronic system, electronic properties 
can be expected to react unusually strongly to slight lattice 
distortions. Such ''overreactions'' of electronic properties have 
indeed been observed in several cuprate compounds.\cite{Axe} 
In particular, a slight orthorhombicity of the lattice structure 
would lead to a relatively strong orthorhombic distortion of the 
Fermi surface. Yamase and Kohno \cite{YK2} invoked this idea to 
explain peculiarities of magnetic excitations in cuprates. 

Large Fermi surface fluctuations could be at least partially
responsible for the non-Fermi liquid behavior observed in cuprates 
at optimal doping. In our model calculation we have obtained a
strongly anisotropic anomalously large decay rate for
single-particle excitations.
Large anisotropic decay rates have been frequently inferred 
from the linewidth observed in photoemission experiments on
optimally doped cuprates.\cite{Val} 
However, recent experiments with higher resolution revealed that 
bilayer splitting of the bands has to be taken into account in 
the data analysis, such that the intrinsic linewidth near the 
van Hove points may be considerably smaller and less anisotropic 
than previously expected.\cite{DSH}
As to the temperature dependence of the decay rate, it is presently 
hard to reliably discriminate a $T$-linear from a $T^{2/3}$ 
behavior of the intrinsic linewidth extracted from the 
experimental data.
Concerning transport, an anisotropic scattering rate with nodes
on the diagonal can very naturally account for the pronounced
anisotropy between intra- and inter-plane mobility of charge
carriers, as pointed out by Ioffe and Millis \cite{IM} in
their phenomenological ''cold spot'' scenario. According to
their idea, the intra-plane transport is dominated by 
quasi-particles with a long life-time near the diagonal of the 
Brillouin zone, while these carriers are not available for
inter-plane transport, since transverse hopping amplitudes
vanish on the diagonal. 
To analyze transport properties near a PI one has to face the 
notorious difficulties of transport theory in a quantum critical 
regime. 
Even for rough estimates of temperature dependences within a
semiclassical Boltzmann approach one would have to compute
subleading corrections to the decay rates, since those limit
the life-time of excitations on the Brillouin zone diagonal, 
and thus the in-plane conductivities.
Furthermore, regular interactions or interactions peaked at
finite q-vectors (leading to ''hot spots'' on the Fermi surface)
should also play a role, and may lead to an interesting but 
complicated interplay with Fermi surface fluctuations.

Since the PI breaks the orientational symmetry of the lattice,
it is natural to consider a possible connection with the tendency 
towards stripe formation, which has been extensively discussed
in the context of cuprate superconductors.\cite{stripes} 
Stripes also break the translation invariance in addition to
orientational symmetry, and their formation requires interactions
with large momentum transfers, such as antiferromagnetic
interactions. The possibility of direct transitions between
states with broken orientational order and stripe states, which
has been envisaged already by Kivelson et al.\cite{KFE}, is an
interesting subject for future studies.

In summary, an electron system close to a PI is characterized 
by a soft Fermi surface, which can react strongly to a slight 
change of the lattice structure, and exhibits strong collective 
Fermi surface fluctuations. 
In two-dimensional systems these fluctuations lead to large 
anisotropic decay rates for single-particle excitations and 
thus to a breakdown of Fermi liquid theory. 
It will be interesting to further explore the consequences
of Pomeranchuk criticality, especially for charge and heat 
transport, for magnetism, and for superconductivity.

\vskip 1cm

\noindent
{\bf Acknowledgements:} \\
We are very grateful to Claudio Castellani for contributing 
several valuable ideas and insights, especially at the initial 
stage of this work.
We also thank B. B\"uchner, B. Keimer, C. Di Castro, G. Khaliullin, 
G. Kotliar, V. Oganesyan, M. Salmhofer, F. Wegner, H. Yamase, 
and R. Zeyher for useful discussions.
This work was initiated at the workshop ''Realistic theories of 
correlated electron materials'' in fall 2002 at ITP-UCSB,
and supported by the DFG-grant Me 1255/6-1.


\vfill\eject


\begin{figure}
\center
\epsfig{file=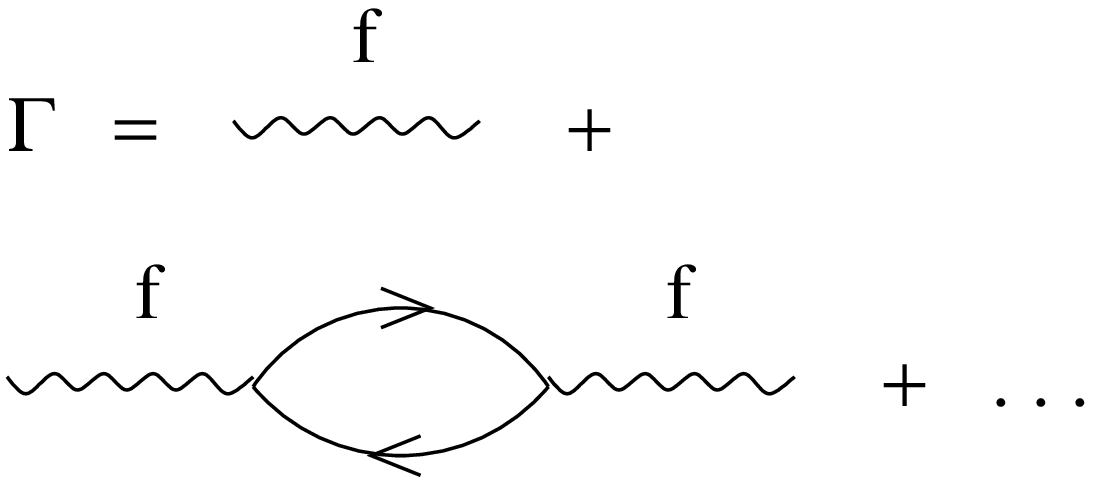,width=8cm}
\vskip 1cm
\caption{Series of bubble chains contributing to the two-particle
 vertex $\Gam$.} 
\end{figure}

\vskip 1cm

\begin{figure}
\center
\epsfig{file=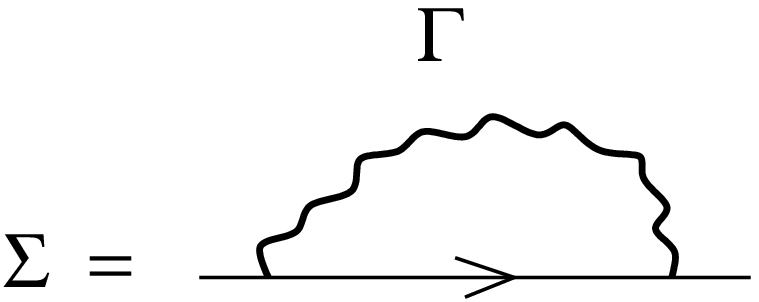,width=5cm}
\vskip 1cm
\caption{Feynman diagram relating the self-energy $\Sg$ to the
 two-particle vertex $\Gam$.} 
\end{figure}


\begin{thebibliography}{99}
\bibitem{fn1} Although the discontinuity of the momentum 
 distribution function is smeared by the gap in a superconductor, 
 a Fermi surface can still be defined as the locus of zeros of the 
 diagonal components of the inverse Nambu propagator in momentum 
 space.
\end{thebibliography}
\end{document}